# Is Metaverse LAND a good investment? It depends on your unit of account!

Voraprapa Nakavachara and Kanis Saengchote*

*Chulalongkorn University*

This version: 7 February 2022

***ABSTRACT***

The Sandbox metaverse LAND non-fungible token (NFT) prices increased by than 300 times (in USD) between December 2019 and January 2022, but when measured in its native utility token (SAND), the increase is only 3 times. Depending on how prices are denominated, investment returns and effective transaction prices vary. We analyze more than 71,000 transactions and find that users are willing to pay 3-4% more when transactions are settled in SAND, and 30% less when settled in wETH (a smart contract version of ETH) when compared to ETH, so unit of account matters. Our results contribute to the discussions of blockchain-based, virtual economy management and the digitalization of money (Brunnermeier et al., 2019).

* Corresponding author. Chulalongkorn Business School, Chulalongkorn University, Phayathai Road, Pathumwan, Bangkok 10330, Thailand. (email: kanis@cbs.chula.ac.th).

0

# 1. Introduction

A metaverse is a digital world where users' avatars can interact and socialize with one another. What distinguishes this new-generation digital world from existing game platforms like Second Life, Roblox, or Minecraft is the fact that it is built on public blockchain. The assets in the metaverses are NFTs (non-fungible tokens – unique digital tokens that are not interchangeable) and can exist on blockchains even outside of the metaverses. In addition, blockchains allow the metaverses to have their own cryptocurrencies and thus their own financial ecosystems, providing a novel setting to test monetary theories. The word metaverse has recently caught public attention since the announcement by Facebook (in October 2021) on changing its name to Meta and transforming into a metaverse company.

The Sandbox was originally launched in 2012 as a game on iOS and Andriod. In 2018, The Sandbox was rebranded and rebuilt on the Ethereum blockchain. It became one of the most popular metaverses and users can purchase LAND and create ASSET to be used in the ecosystem. LAND is an NFT representing a unique digital piece of real estate (identified by coordinates) where 166,464 LANDs (408x408) form a map of The Sandbox metaverse. Each LAND can be bought and sold separately or together as an ESTATE. The LAND/ESTATE presale events (i.e., primary market sales) were held in rounds by The Sandbox. Users may purchase LAND for two reasons: (i) hoping to generate revenue streams, or (ii) hoping to resell at a higher price. In the former case, users can build games and/or interactive experiences on LAND hoping to earn revenues from visitors. In the latter case, they may choose to resell LAND in secondary markets such as OpenSea (a peer-to-peer marketplace for NFTs). Currently, various companies (e.g., Adidas, Atari, PwC Hong Kong, Binance, South China Morning Post) and celebrities (e.g., Snoop Dogg, Pranksy) are reported to have purchased LAND in The Sandbox. The most expensive sale was recorded at USD 4.3 million.[1] SAND is an ERC-20 utility token used as the native currency of The Sandbox metaverse. Users need SAND to transact within the ecosystem, including the primary LAND sales. However, for secondary sales, prices can be freely denominated in any other cryptocurrencies such as ETH, wETH, DAI, USDC, etc.

In this paper, we focus on the returns to LAND investment using real estate analyses such as hedonic pricing regressions and price indices. By examining repeat sales (the exact same bundle of LAND/ESTATE bought or minted and then resold), we can more accurately assess whether purchasing and reselling LAND is a good investment in different denominations. Several papers have documented the economics and returns properties of metaverse real estate (e.g. Goldberg et al., 2021; Dowling, 2022) but base their analyses on USD. Our paper takes a multi-currency approach and sheds light on the monetary implications of a virtual economy built on open blockchain.

# 2. Data and Methodology

We obtain all LAND sale transactions from the Ethereum blockchain and supplement each LAND with data obtained from The Sandbox's API, allowing us to track the contiguity of LAND parcels which can be formed into bundles or ESTATEs.[2] Transactions occurred between December 5, 2019 and January 28, 2022, comprising both primary sales (minted directly from The Sandbox) and secondary sales (via OpenSea). Token price data is obtained from CoinGecko at daily frequency and used to convert transactions to a common denomination. We winsorize prices at thresholds of 0.1% and 99.9% to limit the influence of outliers. There are 116,767 unique transactions when counted at parcel level, but when aggregated into bundles, the number of unique transactions is reduced to 71,205, with lot size ranging between 1 to 576 parcels. LAND age is computed as the difference in days between mint date and transaction date.

---

[1] https://www.wsj.com/articles/metaverse-real-estate-piles-up-record-sales-in-sandbox-and-other-virtual-realms-11638268380, accessed on February 1, 2022.

Table 1 shows the distribution of tokens used for transactions and summary statistics. The time series of LAND transactions are plotted in Figure 1A, which shows that activity increased in late 2021, coinciding with Facebook's announcement of name change. Prior to this announcement, ETH and LAND prices were more highly correlated, as displayed in Figure 2. When SAND appreciated relative to ETH, the share of secondary transactions settled in SAND increased, as evident in Figure 1B.

**Table 1: Summary statistics**
Panel A reports the token denomination of sale by type. Primary sales are LAND minted directly from The Sandbox and secondary sales are auctioned/sold on OpenSea. ETH is the Ethereum blockchain's native coin, which is also used to pay gas, wETH is an ERC-20 token (wrapped) version of ETH, SAND is The Sandbox's ERC-20 utility token that is used to purchase in-game LAND and ASSET introduced in August 2020, and DAI and USDC are ERC-20 token with prices pegged to USD (stablecoins). Prior to SAND's introduction, LAND was minted with DAI. Panel B reports summary statistics of LAND at sale. Repeat sales are defined as transactions involving the bundle of LAND (same lot size and coordinates), which are a subset of secondary sales. Prices are converted to USD using daily prices obtained from CoinGecko. Lot size is the number of LAND parcels that are combined into a bundle. Only transactions involving contiguous parcels are included in the sample. Age is the number of days between the transaction date and the LAND mint date.

Panel A: Denomination of sales by type

| Denomination | Primary | % | Secondary | % | Total | % |
|---|---|---|---|---|---|---|
| ETH | 0 | 0.0% | 30,654 | 74.0% | 30,654 | 43.1% |
| wETH | 0 | 0.0% | 6,753 | 16.3% | 6,753 | 9.5% |
| SAND | 29,587 | 99.4% | 3,848 | 9.3% | 33,435 | 47.0% |
| DAI | 176 | 0.6% | 18 | 0.0% | 194 | 0.3% |
| USDC | 0 | 0.0% | 169 | 0.4% | 169 | 0.2% |
| Total | 29,763 | 100.0% | 41,442 | 100.0% | 71,205 | 100.0% |

---

[2] The Ethereum address for the LAND contract is '0x50f5474724e0ee42d9a4e711ccfb275809fd6d4a'.

Panel B: LAND characteristics at sale

| All transactions: N = 71,205 | | | | Repeat sales only: N = 30,573 | | | |
|---|---|---|---|---|---|---|---|
| | USD price | Lot size | Age (days) | | USD price | Lot size | Age (days) |
| Mean | 5,575.17 | 1.44 | 158.7 | Mean | 8,429.38 | 1.23 | 245.4 |
| StdDev | 13,301.10 | 4.26 | 205.1 | StdDev | 13,589.64 | 1.38 | 191.0 |
| Skewness | 10.38 | 76.44 | 1.05 | Skewness | 8.15 | 7.17 | 0.50 |
| Kurtosis | 193.51 | 9,504.6 | 2.85 | Kurtosis | 107.72 | 70.85 | 2.33 |
| p5 | 39.75 | 1 | 0 | p5 | 331.43 | 1 | 2 |
| p50 | 1,137.54 | 1 | 28 | p50 | 5,569.83 | 1 | 210 |
| p95 | 16,890.18 | 2 | 596 | p95 | 18,758.17 | 1 | 608 |

**Figure 1: LAND transactions**
Weekly LAND transactions and SAND prices are reported in Panel A. Sales include both primary sales (minted directly from The Sandbox) and secondary sales (on OpenSea). For repeat sales, the same bundle of LAND must have been previously bought (or minted) and then sold. SAND is The Sandbox's ERC-20 utility token that is used to purchase in-game LAND and ASSET, so primary sales of LAND are settled in SAND. Panel B reports the weekly share of secondary sales that are settled in SAND against an index of relative price between ETH and SAND. Higher index means SAND price has appreciated relative to ETH. The series in panel A begins at week 49 of 2019 but the series in panel B begins at week 33 of 2020 because SAND was introduced in August 2020.

Panel A: Weekly LAND transactions

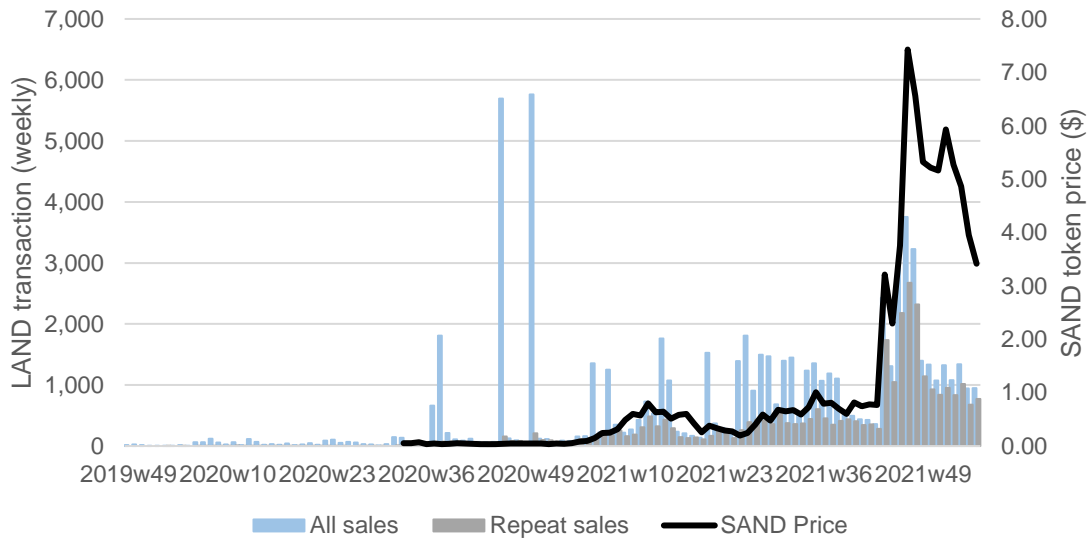

Panel B: Weekly share of secondary LAND transactions settled in SAND versus ETH/SAND index

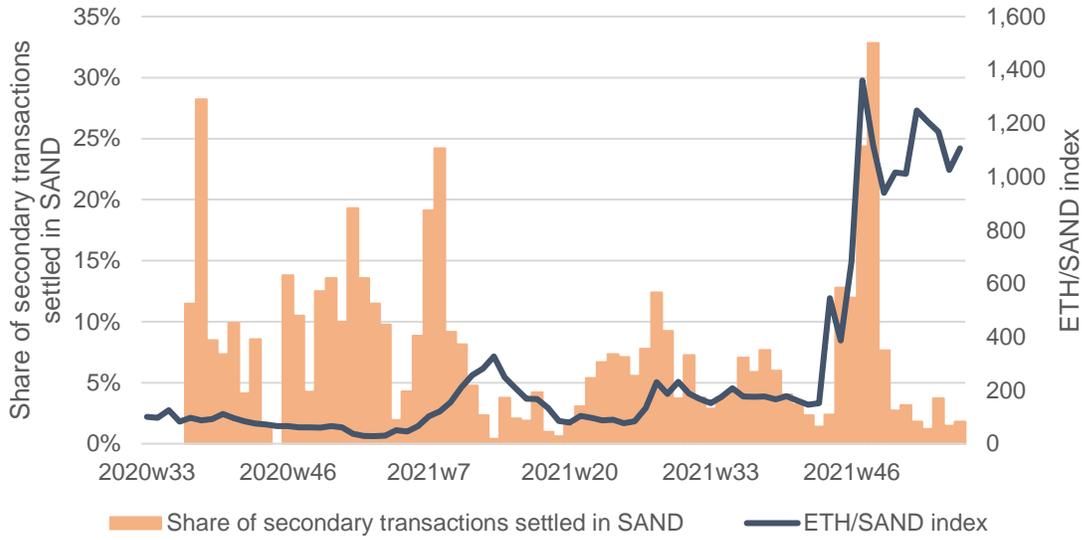

## Figure 2: ETH and SAND prices

Weekly ETH and SAND prices beginning week 33 of 2020 are compared. The correlation coefficient between the two series is 0.6414, but prior to week 44 of 2021 (late October, when Facebook announced it would change its name to Meta), the correlation was 0.8678.

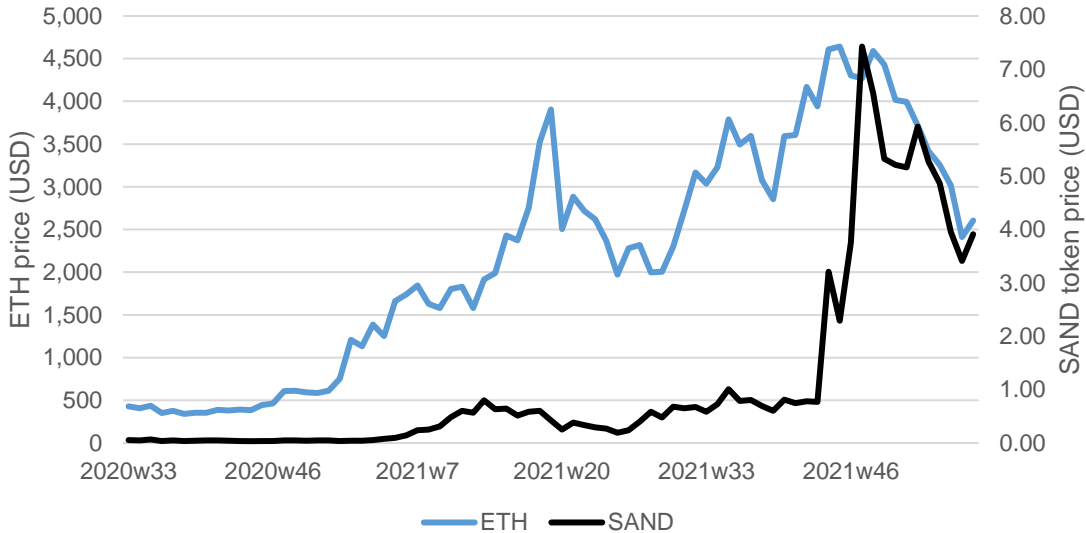

$$p_{it} = \sum_t \delta_t + \sum_j \beta_j x_j + \varepsilon_{it} \qquad (1)$$

To create all-sales price indices, we use the hedonic price index method (Fisher, Geltner and Webb, 1994; Hill, 2013) and regress log prices on controls and indicator variables $\delta_t$ for each week-year as specified in Equation 1. The exponents of $\delta_t$ are the index levels for each period. We compute the indices for three different denominations: USD, ETH and SAND, where ETH is the Ethereum blockchain's native coin (which is also used to pay gas) and SAND is The Sandbox's ERC-20 utility token that is used to purchase in-game LAND and ASSET introduced on 14 August 2020. Users are free to denominate the token of settlement for OpenSea auctions/sales, and the main tokens are ETH, wETH – an ERC-20 token (wrapped) version of ETH, DAI and USDC – ERC-20 tokens with prices pegged to USD (stablecoins), and SAND.[3] We include indicator variables for different settlement tokens with ETH (43.1% of all transactions) as the omitted category. Other control variables include the natural log of lot size, LAND age and indicator variable for primary sale (99.4% occur in SAND).

$$\Delta p_{it} = \sum_t \delta_t + u_{it} \qquad (2)$$

To better measure investment returns, we identify repeat sales by requiring that lot size and coordinates of the LAND bundle are the same when purchase and resold. For example, a sale of a 3x3 bundle that was combined from 9 parcels purchased separately is not a repeat sale, thus repeat sales are a subset of secondary transactions. By using same bundle, time-invariant unobservable characteristics are controlled for, reducing the concern of omitted variable bias. We use the Case and Shiller (1987) method to construct repeat sales indices which requires three steps.[4] First, time indicators are regressed on price changes (Equation 2) to generate residuals. Second, the squared residuals are regressed on holding periods (in weeks) to generate predicted residuals. Third, the predicted residuals are used as weights to re-estimate Equation 2.

## 3. Results

First, we begin with hedonic regressions (Equation 1) in various denominations in Table 2.[5] The adjusted R-squared for USD and ETH prices are very high, suggesting that time trends play an important role in prices. Across all denominations, coefficients on lot size, LAND age and primary sale variables have the same directions and similar magnitudes. Primary sale prices are 33.9% lower on average, reflecting early buyer advantage.

In this paper, we are interested in the influence of denominations, so we turn our attention to the settlement variables. Compared to ETH, transactions settled in wETH and USDC are priced significantly lower (25-30%), while SAND settlements are priced higher (3-4%).[6] On OpenSea, users who make unsolicited offers must use wETH by platform design, but wETH can also be used in auctions. We cannot distinguish between solicited and unsolicited sales, but the large discount is likely a mixture of unsolicited sales and the additional costs of obtaining wETH. Brunnermeier et al. (2019) surmise that "digital money" of the same denomination can be different if users do not view them as perfect substitute – a digital version of Gresham's law.[7] While ETH is pegged to wETH, the friction in obtaining wETH may reduce willingness to pay.

---

[3] ETH and wETH are structurally different because ETH is a native coin which is part the Ethereum blockchain, while wETH is an ERC-20 token created from a wETH smart contract by depositing ETH in exchange for a "wrapped" version wETH. Consequently, wETH can be viewed as a synthetic ETH pegged 1:1. In our sample, the price correlation between ETH and wETH is 0.99997. Most smart contract transactions on blockchain involve ERC-20 tokens, so developers often ask users to use wETH rather than ETH for coding simplicity. The cost of wrapping ETH into wETH is gas, which is not based on the nominal value, so it can be costly to obtain wETH in small amount. For The Sandbox's LAND, users can use ETH, wETH, USDC, DAI, SAND and ATRI (AtariToken, a native coin of the Atari blockchain owned by Atari, a game developer that partners with The Sandbox). ATRI is hardly used on OpenSea so we exclude sales settled in ATRI in our analysis.
[4] The advantage of the Case-Shiller method is that it allows for time-weighting of transactions, so we choose this method. There are other approaches to constructing repeat sales price indices. Nagaraja, Brown and Wachter (2014) compare five methods and conclude that all indices are adequate.
[5] Because SAND was introduced in August 2020, transactions prior to SAND's existence are excluded for analyses denominated in SAND.
[6] DAI is primarily used to mint LAND, so there is little variation left for secondary transactions.

The wedge between SAND and USDC can reflect the buyers' frame of reference. When viewed in USDC (hence USD), LAND may be expensive; but viewed in SAND (which appreciated by almost 200x during the sample period), LAND is cheap. The unit of account effect can influence buyers' willingness to pay, making the effective USDC prices lower and effective SAND prices higher when quoted as USD.

**Table 2: Hedonic regressions of LAND prices**
This table reports the results from the hedonic pricing regressions of the natural log of LAND prices (denominated in various units) on control variables: natural log of lot size, natural log of LAND age, indicator variable for primary sale (minted from The Sandbox) [omitted category is secondary sale], indicator variables for different tokens used to settle the sale [omitted category is ETH], and indicator variables for year-week of transaction. Standard errors are adjusted for heteroskedasticity and reported in parenthesis. Stars correspond to statistical significance level, with *, ** and *** representing 10%, 5% and 1% respectively.

| Denomination | (1) USD | (2) ETH | (3) SAND |
|---|---|---|---|
| ln (lot size) | 0.989*** | 0.988*** | 0.991*** |
|  | (0.01) | (0.01) | (0.01) |
| ln (LAND age) | -0.017*** | -0.016*** | -0.021*** |
|  | (0.00) | (0.00) | (0.00) |
| Primary sale (mint) | -0.339*** | -0.349*** | -0.393*** |
|  | (0.02) | (0.02) | (0.02) |
| Settled in SAND | 0.038*** | 0.031*** | 0.016* |
|  | (0.01) | (0.01) | (0.01) |
| Settled in wETH | -0.304*** | -0.306*** | -0.321*** |
|  | (0.01) | (0.01) | (0.01) |
| Settled in DAI | -0.097 | -0.073 | -1.036 |
|  | (0.13) | (0.13) | (0.68) |
| Settled in USDC | -0.248*** | -0.248*** | -0.246*** |
|  | (0.07) | (0.07) | (0.07) |
| Year-week indicators | Included | Included | Included |
| Observations | 71,205 | 71,205 | 69,682 |
| Adj R-squared | 0.929 | 0.847 | 0.535 |

We construct LAND all-sales price indices from $\delta_t$ and report them alongside SAND price in Figure 3. The correlation between the USD index and SAND is 0.9779, while the correlation to the SAND index is -0.0291. When viewed in USD, LAND prices reached the maximum of 302x, but only 11.6x in ETH and 3.0x in SAND. When the SAND price effect is purged from LAND index construction, price appreciation is much closer to traditional real estate.

---

[7] Gresham's law refers to a situation where two types of money with fixed exchange ratio are not viewed as perfect substitute. The classic example is gold and silver coins, where gold coins (preferred, or "good" money) are hoarded and silver coins ("bad" money) spent, potentially leading to deviation from the fixed exchange rate.

**Figure 3: LAND all-sales price indices**

Weekly LAND all-sales price indices are constructed using the hedonic price index method (Fisher, Geltner and Webb, 1994; Hill, 2013). Both primary sales (minted directly from The Sandbox) and secondary sales (on OpenSea) are included in the index construction. Transaction prices for the indices are denominated in USD (panel A), ETH (panel B) and SAND (panel C). ETH is the Ethereum blockchain's native coin which is also used to pay gas, and SAND is The Sandbox's ERC-20 utility token that is used to purchase in-game LAND and ASSET. Indices begin at week 49 of 2019 for USD and ETH but at week 33 of 2020 for SAND because the token was introduced in August 2020. For each index, the correlation coefficient between the index level and SAND token price is reported.

Panel A: LAND all-sales price index denominated in USD (correlation = 0.9779)

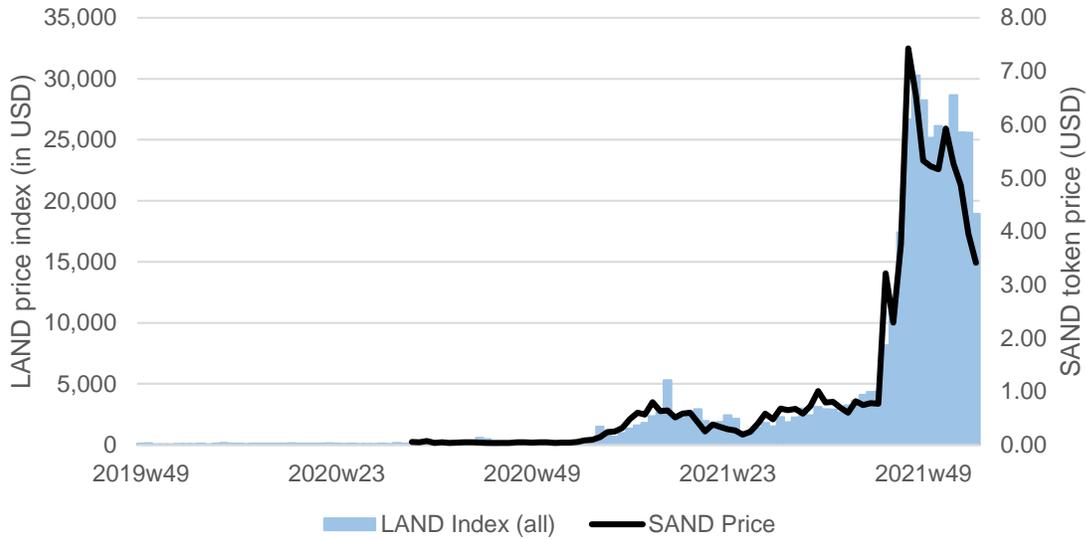

Panel B: LAND all-sales price index denominated in ETH (correlation = 0.8141)

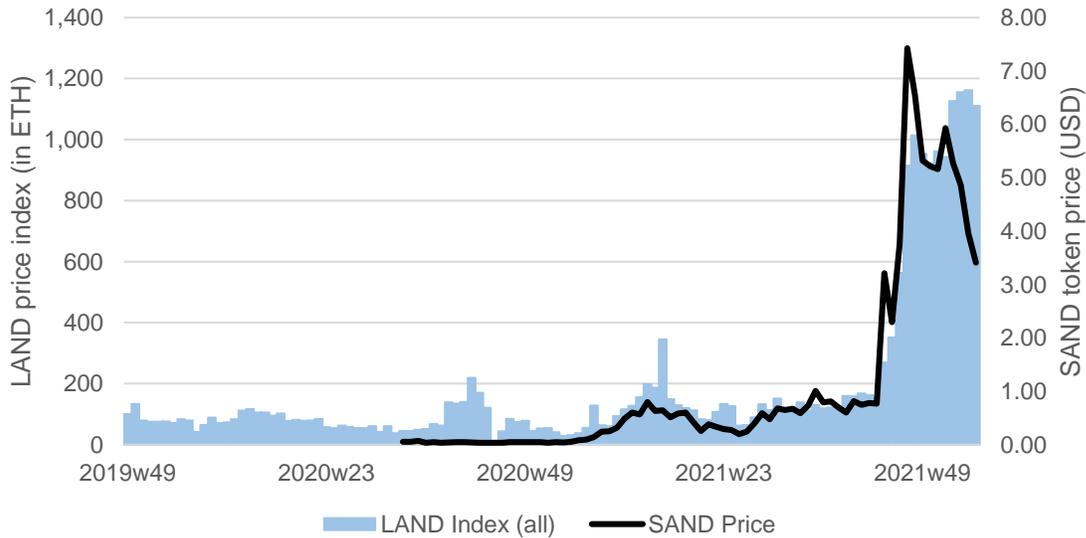

Panel C: LAND all-sales price index denominated in SAND (correlation = -0.0291)

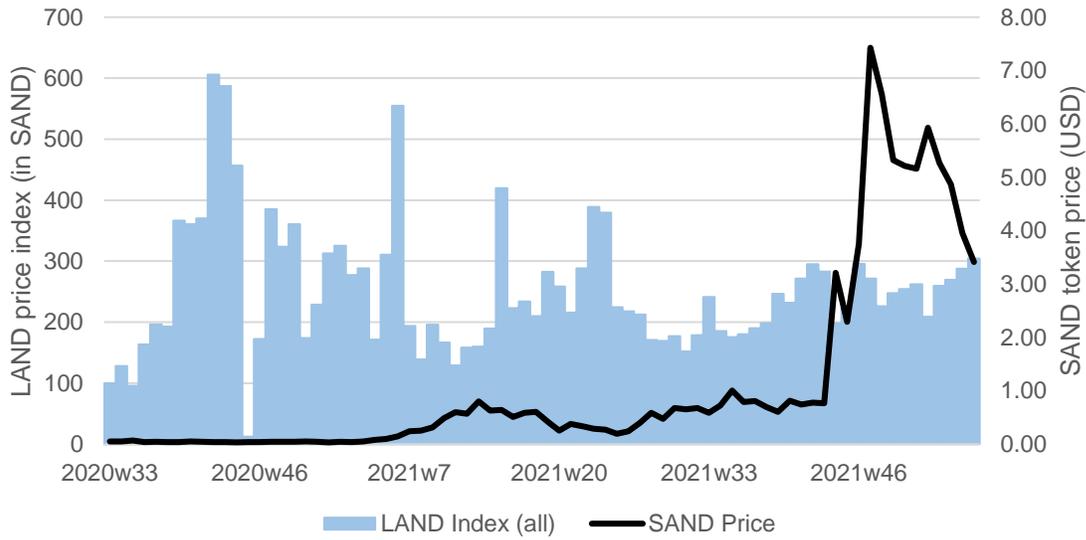

All-sales indices are good for tracking general prices but do not directly assess investment returns, since weekly sales can involve new real estate with different (unobservable) attributes. We use repeat sales on the same parcel and compute multiple on invested capital (MOIC) in USD, ETH and SAND, which are reported in Table 3. The distributions of MOICs are highly positively skewed, especially for USD. Figure 4 plots the histograms, capping MOIC at 20x for ease of comparability. The lower average returns in SAND are consistent with the moderated SAND index observed in Figure 3C.

**Table 3: Returns statistics of repeat sales**
Returns on repeat sales are presented as multiple on invested capital (MOIC), which is computed as the current transaction price divided by the previous transaction price. Consequently, MOIC of 2x means 100% return on invested capital. Prices are converted to USD, ETH and SAND using daily prices obtained from CoinGecko.

|  | USD | ETH | SAND |
|---|---|---|---|
| Num Trans | 30,573 | 30,573 | 29,809 |
| Mean | 25.03 | 5.18 | 1.59 |
| StdDev | 67.67 | 8.89 | 1.19 |
| Skewness | 0.52 | 0.44 | 0.31 |
| Kurtosis | 1.33 | 1.27 | 1.01 |
| p5 | 2.54 | 2.00 | 1.34 |
| p50 | 10.89 | 4.76 | 1.91 |
| p95 | 342.38 | 41.94 | 4.95 |

### Figure 4: Multiple on invested capital of LAND sales

Multiple on invested capital (MOIC) is computed as the current transaction price divided by the previous transaction price. Displayed MOICs are capped at 20x for ease of comparability. Prices are converted to USD, ETH and SAND using daily prices obtained from CoinGecko.

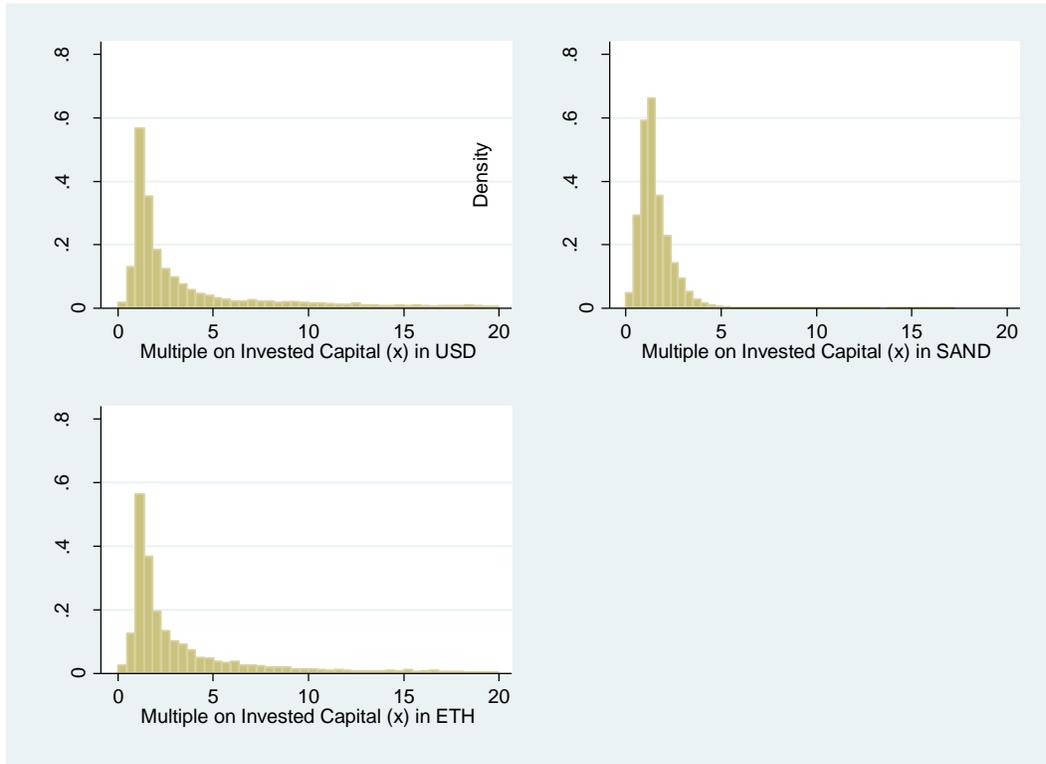

### Figure 5: LAND price indices from repeat sales

Weekly LAND price indices are constructed from repeat sales transactions using the Case and Shiller (1987) method. Only secondary sales (on OpenSea) that have identifiable previous transactions are included in the index construction. Transaction prices for the indices are denominated in USD (panel A), ETH (panel B) and SAND (panel C). ETH is the Ethereum blockchain's native coin which is also used to pay gas, and SAND is The Sandbox's ERC-20 utility token that is used to purchase in-game LAND and ASSET. Indices begin at week 33 of 2020 to allow for sufficient repeat sales in each week and to coincide with the introduction of SAND in August 2020. For each denomination, the correlation coefficient between the repeat sales index level and the all-sales index level are reported. For Panel B and C, the scales are the same for all-sales and repeat sales indices.

Panel A: LAND repeat sales price index denominated in USD (correlation = 0.7828)

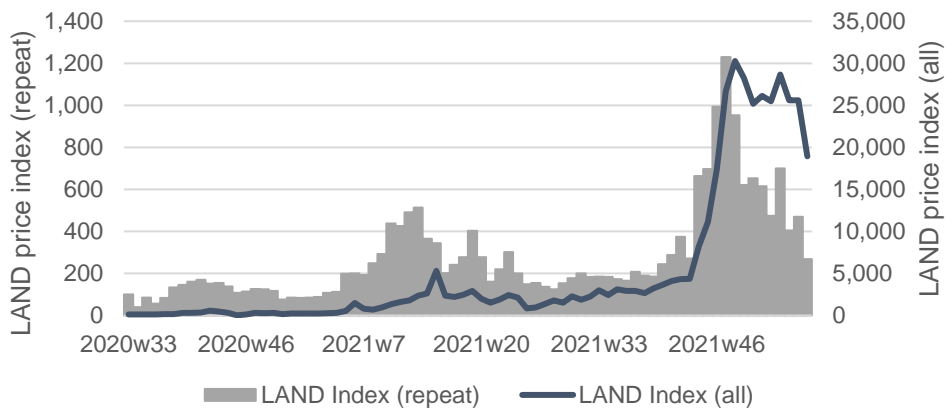

Panel B: LAND repeat sales price index denominated in ETH (correlation = 0.8332)

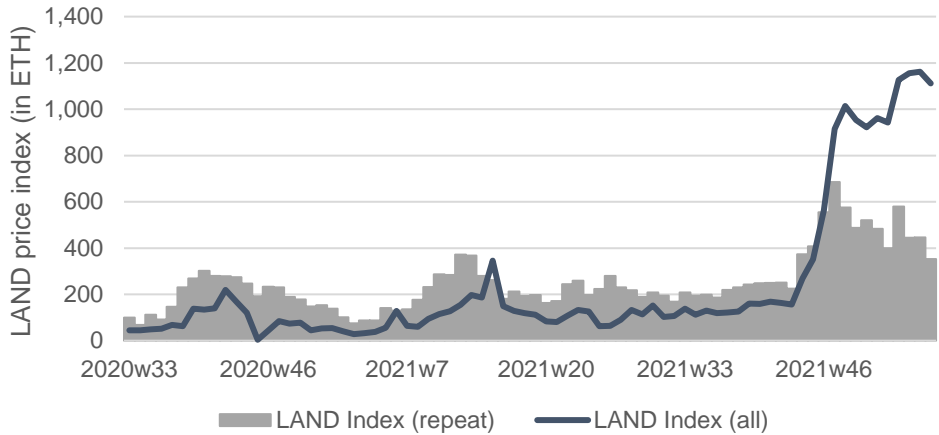

Panel C: LAND repeat sales price index denominated in SAND (correlation = 0.1268)

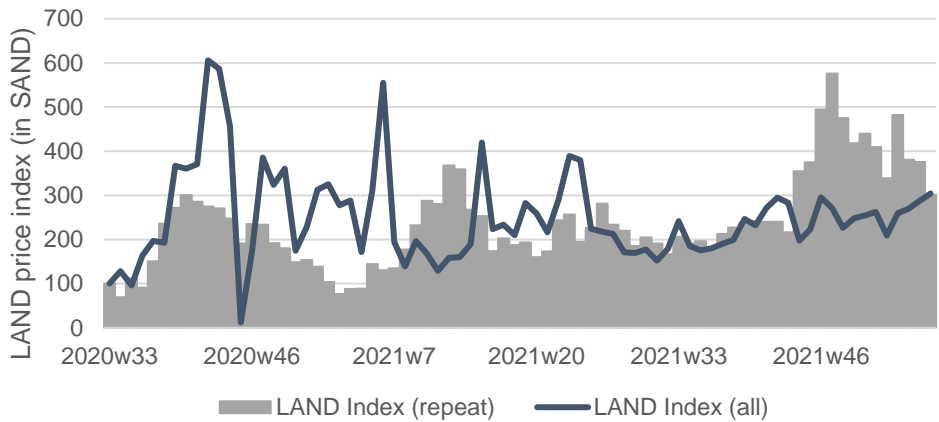

The Case-Shiller indices are reported in Figure 5 along with all-sales indices from Figure 3. The levels for USD and ETH indices are much lower, with the maximum levels at 12.3x and 6.9x respectively, suggesting that all-sales indices could be driven by composition change. The correlations between the two types of indices for each denomination are also high, at 0.7828 and 0.8332. The fall in USD indices began in December 2021, coinciding with a fall in both ETH and SAND prices. For SAND indices, the trends are slightly different, and the correlation is much lower at 0.1268. From these pictures, one may reach different conclusions about LAND's performance under different denominations.

Combined with earlier results where willingness to pay is related to denomination, our findings have important implications for management of open, virtual economy: unit of account matters. Viewed as The Sandbox's native, LAND price did not appreciate that much and perhaps could reflect the anticipated productivity of LAND denominated in SAND. But viewed as a "foreigner" (in USD or even ETH), LAND became much more costly by the end of 2021. Users who were able to obtain SAND at lower bases may behave differently, highlighting the behavioral influence of unit of account.

Metaverses are virtual economies whose ecosystems require careful management. Virtual world developers often carefully design and closely monitor the balance of the economy, and the introduction of unintended wealth such as real-money trading (the practice of purchasing in-game items with real-world monetary side agreements) can lead to price inflation and upset the economic incentives and users' experiences, as discussed by Knowles and Castranova (2016). In this regard, The Sandbox can be considered an open economy with no capital control and fully flexible exchange rate, as the permissionless convertibility of blockchain tokens effectively allows perfect capital mobility. Under the Mundell-Fleming view of the "Impossible Trinity" (formalized by Dornbusch, 1976), this virtual economy has no monetary sovereignty and has a fully liberalized financial system, which can be hard to manage. Also, the inability to meaningfully restrict the medium of exchange for secondary transactions means virtual economies' own currencies can easily be supplanted (or "dollarized"), as illustrated by The Sandbox's OpenSea auctions/sales.

## 4.     Conclusion

In this paper, we document LAND price appreciation in The Sandbox metaverse using methods employed for real estate analyses. We find that Case-Shiller repeat sales indices show more moderated appreciation than all-sales indices, and show even less appreciation when denominated in SAND. For a virtual economy native who views activities in SAND, LAND price did not appreciate that much. We also find that transactions settled in different denominations can influence willingness to pay, with wETH-settled LAND priced 30% lower, while SAND-settled LAND priced 3-4% higher, highlighting the psychological effect of "native" unit of account.

In the context of digitalization of money of Brunnermeier et al. (2019), the wETH-ETH wedge is an example of "digital Gresham's law", and the ability to freely form side transactions outside the ecosystem's control in any denomination highlights the ease of "digital dollarization" in an open, permissionless world enabled by public blockchains. These findings are also relevant for real economies. Financial liberalization has its benefits but also comes with concerns such as instability (Stiglitz, 2000), and current academic evidence is mixed (Kose et al., 2009). We hope that our findings contribute to the discussions on the implications of a virtual economy and digital money built on public blockchains.